# *Real-time* observation of non-equilibrium liquid condensate confined at tensile crack tips in oxide glasses

Lothar Wondraczek[1,*], Matteo Ciccotti[2], Anne Dittmar[1], Carina Oelgardt[1], Fabrice Célarié[1], Christian Marlière[2,#]

[1] *Institute of Non-Metallic Materials, Clausthal University of Technology, Germany*

[2] *Laboratoire des Colloïdes, Verres et Nanomatériaux, University of Montpellier II, France*

[#] marliere@univ-montp2.fr

[*] lothar.wondraczek@tu-clausthal.de

## Abstract

Since crack propagation in oxide materials at low crack velocities is partly determined by chemical corrosion, proper knowledge of the crack tip chemistry is crucial for understanding fracture in these materials. Such knowledge can be obtained only from *in situ* studies because the processes that occur in the highly confined environment of the crack tip are very different from those that take place at free surfaces, or that can be traced *post mortem*. We report the occurrence of hydrous liquid condensate between the two fracture surfaces in the vicinity of the tip of tensile cracks in silica. Observations are performed in *real-time* by means of atomic force microscopy (AFM) at continuously controlled crack velocities in the regime of stress corrosion. Condensate formation and changes in extent and shape are demonstrated for a wide range of macroscopic humidity at different crack speeds. Its liquid character is confirmed by the study of AFM phase-contrast data. It is believed that this evidence of a nanoscale liquid hydrous phase at the crack tip will enable novel insights in the chemistry of failure of oxide materials.

# 1. Introduction

Crack propagation phenomena are still one of the prominent challenges in fracture mechanics of brittle glasses.[1,2] Concerning oxide glasses at low crack speed, fracture is classically related to stress corrosion: a phenomenon that is generally understood as an accelerated corrosion reaction due to stress concentration at the crack tip.[3] Water is known to be the crucial reactant in this reaction.[4,5] More recently, Tomozawa et al. showed that water diffusion in silica is enhanced by stresses at the crack tip and, consequently, affects the local mechanical properties at the crack tip.[6] As hypothetically mentioned by Wiederhorn, high gaseous humidity environments should result in crack tips that are wetted with *liquid* water, or more likely a reaction product between this condensate and the glass.[7] Thus, hydrous condensates at the tips of cracks in air have been postulated, but never to our knowledge reported in the literature. The chemical character of the crack tip is controlled by the chemical composition of the glass fracture surfaces and reactions with water. In silica glass, e.g., that should result in an acidic environment at the crack tip. In glasses containing mobile alkali ions, ion exchange and corrosion processes accordingly should create a highly basic solution at the crack tip.[7] Hence, the presence of liquid water near the crack tip is crucial for ion exchange processes. However, differences are expected when working i) with the sample immerged in liquid water or ii) in the case of liquid water confined in a, what we call, condensate region: This region is localized between the two crack surfaces at small distance from the crack tip where their distance is of the order of typical interatomic length. On the other hand, the crack tip chemistry is still not evident, since prior studies were performed at free surfaces, rather then in very confined conditions. In order to better understand interdiffusion/corrosion processes that occur in the vicinity of the crack tip, it is therefore of fundamental importance to experimentally verify the presence of liquid water at the tip of a crack propagating in a gaseous atmosphere, and of the conditions for its presence.

## 2. Experimental

In the present study, we used the double cleavage drilled compression (DCDC) technique to create tensile cracks with highly controllable crack speeds in silica glass.[8, 9] Generation and propagation of the crack were performed at 22.0±0.5 °C in a carefully controlled atmosphere of nitrogen, with a relative macroscopic humidity (rh) that was varied between ~ 1 % and 80 % (± 2 %). As sample material, type III silica glass (Suprasil 311, Heraeus, Germany) with a bulk $OH^-$ - content of 200 ppm and soda lime silica glass, respectively, were employed. Samples were cut into 4x4x40 mm³ rods and mechanically polished with $CeO_2$ to a RMS roughness of 0.25 nm for an area of 10x10μm². The crack motion was observed for velocities from $10^{-9}$ m/s to ~$10^{-12}$ m/s by monitoring the line of intersection of the moving crack with the polished glass surface via atomic force microscopy (D3100, Veeco, NY, USA) in a high-amplitude resonant mode (*tapping mode*). Optical microscopy was utilized to record higher crack velocities. For an applied stress $\sigma$, the stress intensity factor $K_I$ in a DCDC experiment is given by $K_I = \sigma a^{1/2}/(0.375c/a+2)$, where $a$ is the radius of the hole in the DCDC specimen and $c$ is the length of the crack.[8] Hence, $K_I$ and crack velocity, respectively, are decreasing with increasing crack length. This makes the DCDC experiment superior to other fracture techniques since control over the crack motion can be maintained very easily. Locations of AFM and optical microscope, and sample orientation are schematically shown in Figure 1.

## 3. Results and Discussion

Operating the AFM in tapping mode yields information on both sample topography and, by phase imaging, local mechanical properties. Considering hydrous condensates in the vicinity of the tip of a surface crack in inorganic glass, those should generate a high phase contrast[10,11]. Figure 2 summarises topography and phase information on the propagation of the tip of a tensile crack in silica at moderate humidity. The velocity of the crack was, in this case, in the range of 0.1 nm/s at a macroscopic relative humidity (rh) of 45 % (22.5 °C) and the corresponding stress intensity factor $K_I$ of 0.37 N/m$^{3/2}$. The crack tip topography indicates three different zones: The very tip of the crack, where damage cavity formation[12] can be observed. This is followed by what looks like a ridge. At the considered conditions, this ridge is ~ 70 nm in length, 10-30 nm in width, and some nm in height. Length increases with increasing humidity and decreasing crack speed. The ridge is finally followed by the two fracture edges at the crack opening. It must be noted that the height contrast on the ridge is small, meaning that its impact on eventual spurious changes in phase is negligible. On the other hand, phase images (Fig. 2) reveal a strong negative contrast between the region of the ridge and the surrounding glass surface. This is associated to strongly increasing dissipation in the ridge zone. The extent of dissipation on a silica surface in AFM tapping mode can be related to the presence and growth of liquid water films.[13]

Furthermore, after exposure to more elevated humidity conditions (rh > 60 %), we observed that the region of high phase contrast was found to be unstable in terms of AFM response (Figure 3). Trace-retrace curves (scanning the AFM tip along one single line –slow scan axis- perpendicular to the crack consecutively from left to right and back again, and comparing both scans) almost perfectly matched for lines that do not cross the ridge zone (Fig. 3a). On the contrary, phase and height signals for lines crossing the ridge zone revealed unstable behaviour when the AFM tip was tracing and retracing (Fig.3b). The instability (reflected by a change in the AFM working point) is triggered by crossing the ridge-zone of the crack tip (corresponding to the peak at ~ 300 nm in Fig. 3b) and persists for a certain distance, eventually snapping back to the original working point (e.g. at ~ 450 nm during tracing in Fig. 3b). The presence of a second, metastable, working point indicates an abrupt change in the tip-surface interactions. In conjunction with the strong negative phase contrast (Fig. 2), the second zone is consequently believed to be liquid.

The geometrical size of the condensate zone was observed to increase in size up to 180 nm by increasing rh up to 80 %. After decreasing the rh again down to 5 %, however, it took up to 5 days, in our experimental conditions, for the condensate to start reducing its size, what can be related to kinetic reasons.

Since with the AFM, only surface of a sample can be considered, optical microscopy was used to image the crack front in the bulk of the silica glass specimen, even though at a lower spatial resolution. Thereby, working in an interferometric reflection mode yields a sharp contrast between connected parts of the two fracture surface (such as unbroken parts of the sample and water bridges) and disconnected parts (fully opened crack regions), respectively corresponding to black (white respectively) parts of the image, when looking perpendicular to the direction of crack propagation. Figure 4 shows the position of the thus determined "crack front" at different times in a soda lime silica glass at ~ 41 % rh (Note that for these studies, soda lime silica was employed rather than silica glass. Condensate formation, in this case, is strongly influenced by the presence of alkali ions and occurs in the µm-range, what makes it easily observable by optical microscopy). Although the direction of crack propagation is from left to right, the "crack front" is moving backwards during time at very low crack speeds. Additionally, the optical contrast at the "crack front" is increasing. This is associated with the formation of a water meniscus (water front) at the real crack front –almost undetectable form optical analysis) which results in an optical contrast that is more abrupt at the water/air interface. These observations were performed after increasing the humidity from a dry state with the crack already running. This means that during the first days of observation, the condensate zone is building up to an optically visible size of a few µm, growing faster, and in a reverse direction, than the crack propagates. The speed of condensate formation was estimated to be in the range of $2*10^{-10}$ m/s for soda lime silica glass. From these investigations, it is concluded that condensate formation occurs alongside the whole crack front, i.e. from bulk to surface, rather than only at the surface of the sample. This effect must be taken into account when determining the crack velocity in the low speed regime ( $< 10^{-9}$ m/s) from optical data.

For elevated humidities, capillary condensation has to be considered when discussing the origin of liquids at the crack tip. According to Crichton and Tomozawa[14] who described capillary condensation by the well known Kelvin equation, at ambient pressure and room temperature condensation is expected at crack tips in silica (for a crack tip radius of 1.5 nm[15]) at rh ~ 50 % (water vapor pressure ~ 1.36 kPa). Even though the condensation process alongside a running crack is out of equilibrium, at low crack speeds this is believed to be an useful approximation. However, it is known that the Kelvin equation is not strictly correct in very confined conditions as they are present at the crack tip and inside the crack opening (e.g. [16]). In this context, the capillary pressure is expected to be strongly related to, e.g., roughness and electric charges. Furthermore, liquids found at the crack tip will not be pure water, but solutions of water, alkali and/or silica and thus, the chemical activity of the solute has to be considered in the Kelvin equation. Since liquid condensate is demonstrated to occur at crack tips already at significantly lower relative humidities than those predicted from the classical Kelvin equation (and water films are known to form on oxide surfaces at water vapor pressures well below 1 Pa), all those effects obviously stabilize the condensate, and only the strong increase of its amount at around 30-50 % rh is related to "common" capillary condensation.

## 4. Conclusions

The formation of liquid condensate at the tip of a sub-critical crack in silica glass was observed in humid nitrogen atmosphere by AFM *in situ* measurements and correlated to optical observation all along the crack line. The characteristic size is in the range of 100 nm, and increasing with increasing humidity and decreasing velocity.

Although water-diffusion processes through the bulk glass structure occur at the crack tip, they can presently not be imaged by tapping mode AFM because their contribution to dissipation is too low compared to that caused by the liquid condensate or as well by the liquid layer on the glass surface. Accordingly, no contrast in AFM phase images can be observed in the vicinity of the crack opening after the condensate zone passed. The condensate region is moving along with the crack advance. This can be explained by its liquid character and the role of surface tension along with the creation of new surfaces in highly confined conditions. The longitudinal speed of condensate formation in silica glass is in the range of $10^{-10}$ m/s at the considered conditions. This has to be taken into account when determining the crack speed from optical data at similar or lower absolute crack velocities. Regarding the formation of cavities at the crack tip[9], it is presently not clear whether they are formed in the condensate region.

## Acknowledgements

Financial support by the German Academic Exchange Service, Germany, and Égide, France, (DAAD-PROCOPE) under contract No. D/0333567 is gratefully acknoledged.

**Figure Captions**

**Fig. 1.** Schematic of the DCDC specimen. Areas that were observed by AFM and optical microscopy (OM), respectively, are highligthed.

**Fig. 2.** Sequential AFM images of the intersection of a moving crack front with the surface of a silica DCDC specimen at room temperature. The crack propagates from the upper left to the lower right at a speed of $10^{-10}$ m/s. Relative humidity is 45 % ±2 % in nitrogen atmosphere. Image size is 50x50 nm$^2$ (left: height contrast, height scale is 2 nm; right: phase contrast, phase scale is 40°).

**Fig. 3.** Trace-retrace height and phase profiles of the surface of a silica DCDC specimen perpendicular to the direction of the moving crack. Profiles were derived from left-to-right and right-to-left AFM linescans at a position where the slow-speed scan line does not cross the ridge zone (Fig. 3.a) or does cross it (Fig. 3.b). See text for details.

**Fig. 4.** Position of a crack front in soda lime silica as seen by optical microscopy. Inset: optical micrograph of the crack front at different times seen in interferometric reflectance mode. Viewing angle is perpendicular to the direction of crack propagation. The broken glass appears white (left) because of reflection at the fracture surfaces, while the unbroken glass appears black (right). Small black spots on the broken side are due to bridges of condensated water near the crack front. The crack is propagating from left to right. We observe that the optical contrast accessible by microscopy is due to water front moving in *reverse* direction. Relative humidity was 41 % ± 2 %.

Figure 1
(Wondraczek, Marlière, *et al.*)

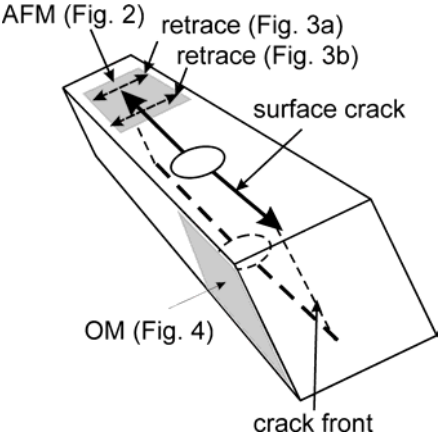

Figure 2
(Wondraczek, Marlière, *et al.*)

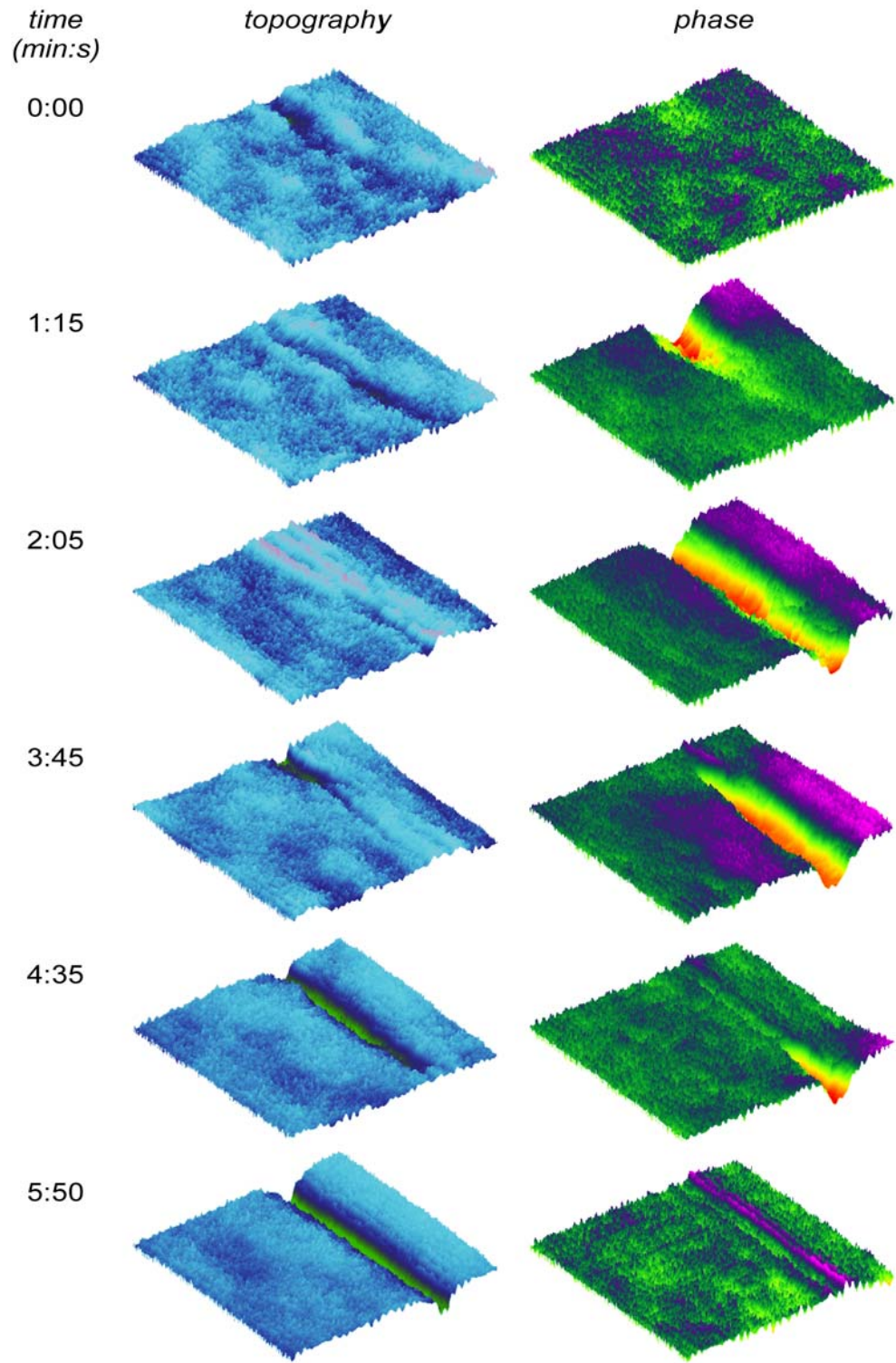

Figure 3.
(Wondraczek, Marlière, *et al.*)

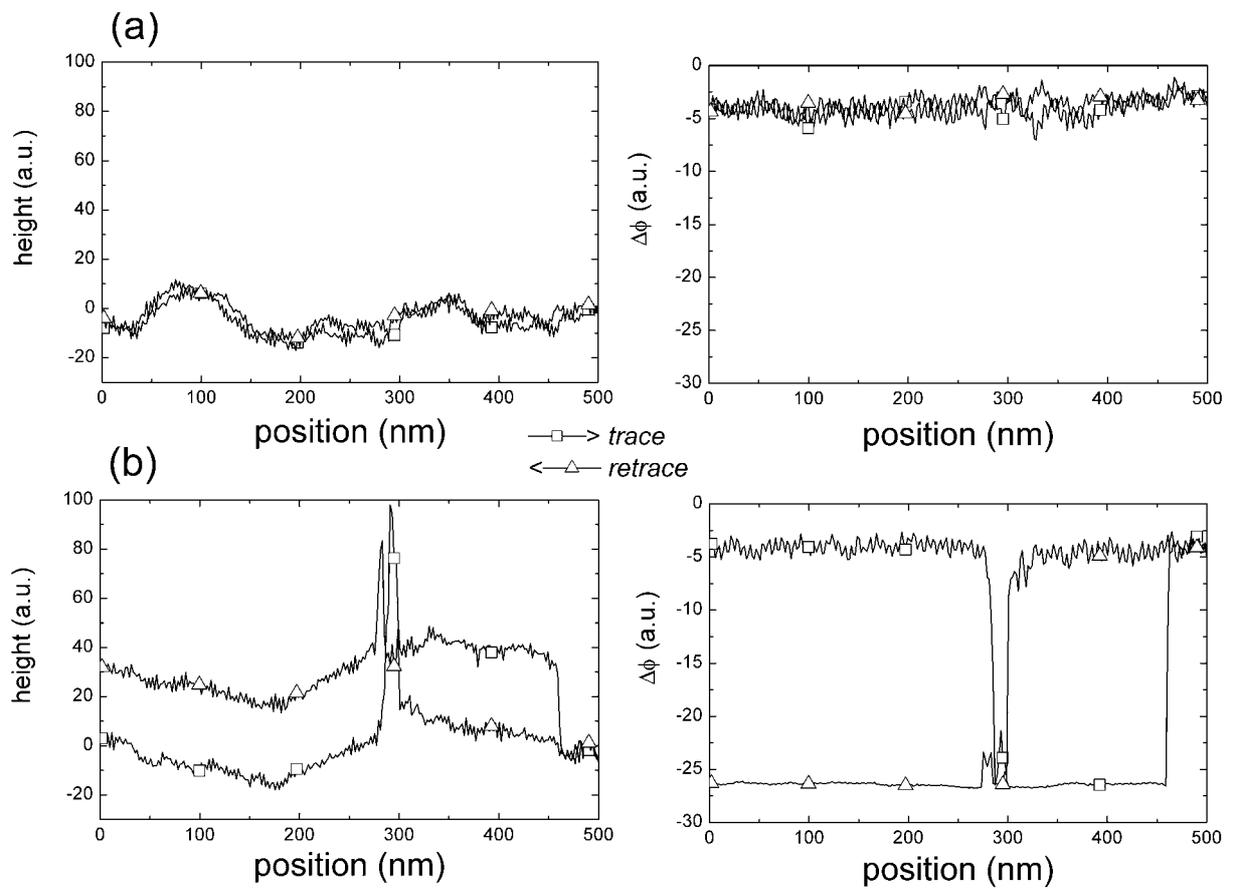

Figure 4.
(Wondraczek, Marlière, *et al.*)

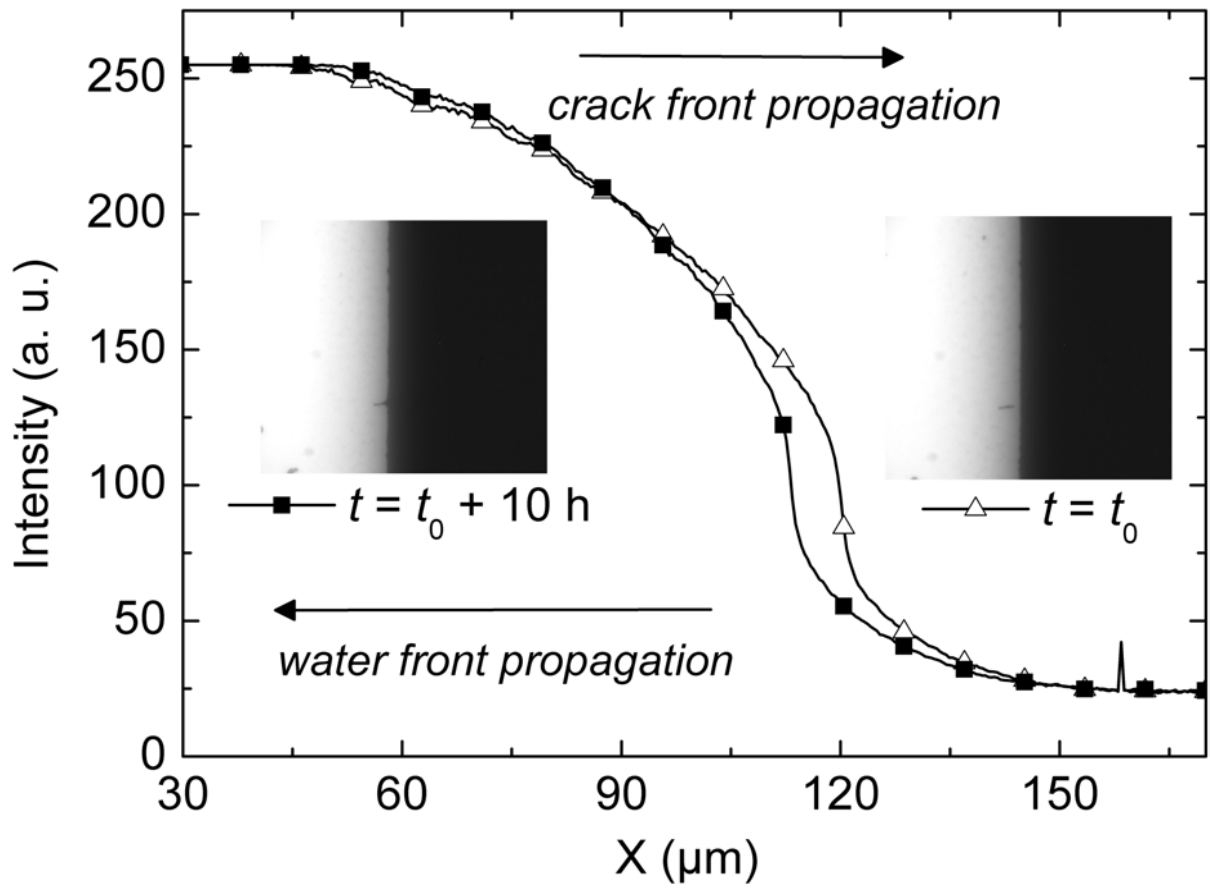